\documentclass{PoS}
\def\GeV{{\rm GeV}}
\def\TeV{{\rm TeV}}

\title{Summary of Structure Functions and PDFs Working Group}

\ShortTitle{WG1 summary}

\author{R.~McNulty \\
        School of Physics, University College Dublin,\\
        Dublin 4, Ireland\\
        E-mail: \email{ronan.mcnulty@ucd.ie}}

\author{\speaker{R.\ S.~Thorne} \\
        Department of Physics and Astronomy, \\
        University College London, WC1E 6BT, UK\\
        E-mail: \email{robert.thorne@ucl.ac.uk}}

\author{\speaker{K.~Wichmann} \\
        Deutsches Elektronen Synchrotron DESY, Hamburg, Germany\\
        E-mail: \email{kklimek@mail.desy.de}}

\abstract{
This summary presents personal highlights from the
Structure Functions and PDFs Working Group (WG1).}

\FullConference{XXIV International Workshop on Deep-Inelastic Scattering and Related Subjects\\
		11-15 April, 2016\\
		DESY Hamburg, Germany}

\begin{document}

\section{Introduction}

The Structure Functions and PDFs Working Group (WG1) consisted of 43 presentations spread over 10 sessions, three of which were held jointly with 
QCD and Hadronic Final States (WG2),
Electroweak Physics and Beyond the Standard Model (WG3),
and Future Experiments (WG7).
This summary presents personal highlights from these sessions, with the exception of the joint session
with WG7 that is presented elsewhere.

\section{PDF Fits including HERA combined data}

The publication of the HERA combined inclusive data~\cite{Abramowicz:2015mha} 
is a legacy document that is at the core of all PDF extractions.
A number of groups considered the impact of this data on 
different PDF sets.
 
\begin{figure}
\begin{center}
\includegraphics[width=13cm]{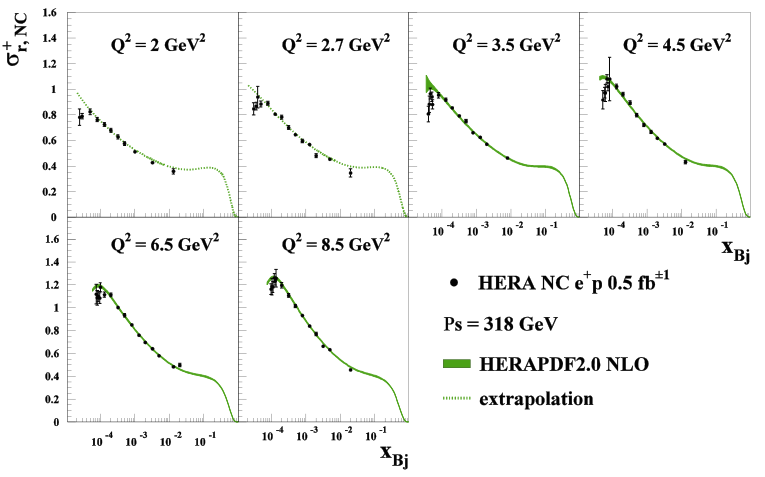}
\caption{HERA neutral current data with HERAPDF2.0 NNLO fit superimposed \cite{Abramowicz:2015mha}.}
\label{fig:coopersarkar}
\end{center}
\end{figure}

The HERAPDF2.0 set~\cite{Abramowicz:2015mha}, derived only from the HERA data,
describes the neutral data well for $Q^2>2$ GeV$^2$ but there are discrepancies at low-$x$ and low-$Q^2$ (Fig.~\ref{fig:coopersarkar}).  A possible resolution is provided by the inclusion of higher
twist effects.~\cite{coopersarkar,Harland-Lang:2016yfn}.
Another approach using the Bartels-Golec-Kowalski dipole model was shown to describe the HERA data
very well, but only if sizeable saturation effects are included~\cite{luszczak}.

\begin{figure}
\begin{center}
\includegraphics[width=7.5cm]{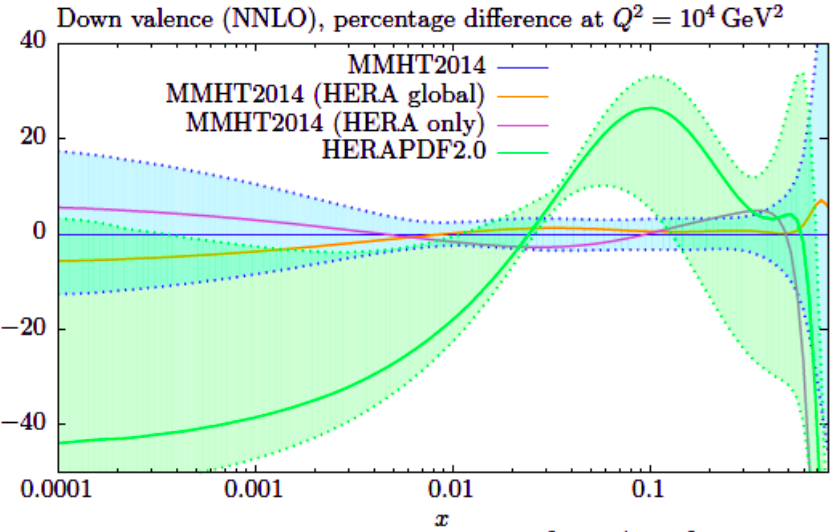}
\includegraphics[width=7.5cm]{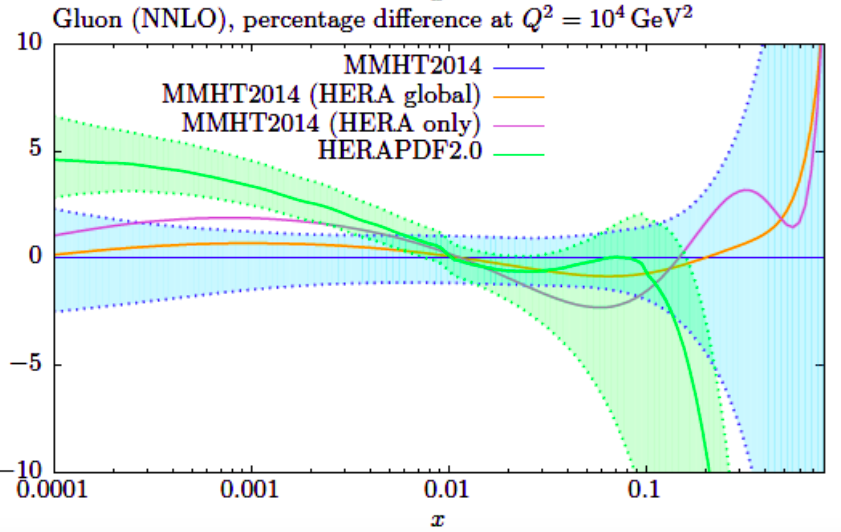}
\caption{Down valence (left) and gluon PDF (right) compared to MMHT2014 for: a global fit including the
new HERA combined data; a fit just using the HERA data; HERAPDF2.0~\cite{Harland-Lang:2016yfn} .}
\label{fig:thorne}
\end{center}
\end{figure}

A fit within the MMHT framework~\cite{Harland-Lang:2016yfn} was shown to be in good agreement 
with the MMHT2014  
PDFs that used previous HERA cross-section data.
There is a very small change in central values and 
the uncertainties reduce a little -- at most by 20\% e.g. the cross-section for $gg\rightarrow H$ at
14 TeV changes from $47.69^{+0.63}_{-0.88}$ to $47.75^{+0.59}_{-0.72}$ pb.
A comparison with HERAPDF2.0 showed broad agreement but there were some marked differences
e.g. in the down valence quark and gluon at high $x$ as shown in Fig.~\ref{fig:thorne}.

\begin{figure}
\begin{center}
\includegraphics[width=14cm]{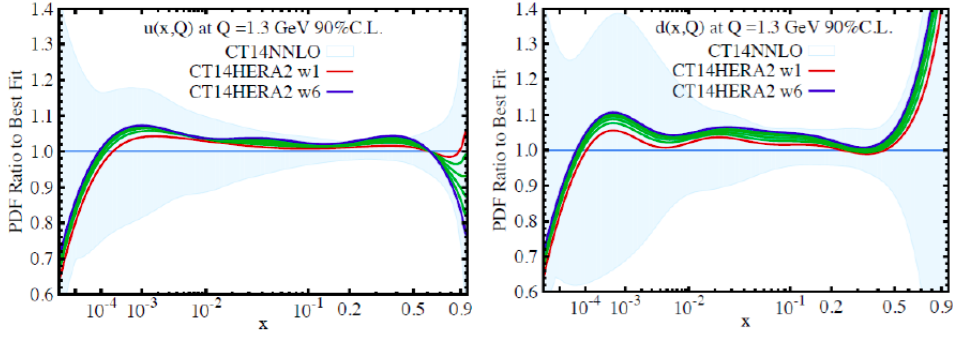}
\caption{Up and down valence PDF compared to CT14NNLO including HERA data with different weights \cite{Schmidt}.}
\label{fig:schmidt}
\end{center}
\end{figure}

Both MMHT and CT groups find
tension between the HERA combined $e^-p$ charged current data
and other data sets.  In a CT14-like fit \cite{Schmidt}
a shift in the up quark near $x=0.3$ is seen if high 
weight is given to the HERA data (see Fig.~\ref{fig:schmidt}).  
It will be interesting to look for this effect in related observables
e.g. high rapidity $W^+$ production at the LHC.

\section{Analysis and comparison of global PDF sets}

There was a lively discussion on how to synthesise the results of the various PDF global fits
in order to obtain a theoretical uncertainty on an observable.
An extreme example was provided by the $H+t\bar t$ production cross-section which changes by 13\%
depending on which PDF set is used.
Various approaches were discussed as expounded in \cite{Butterworth:2015oua} and \cite{Accardi:2016ndt}.

\begin{figure}
\begin{center}
\includegraphics[width=6cm]{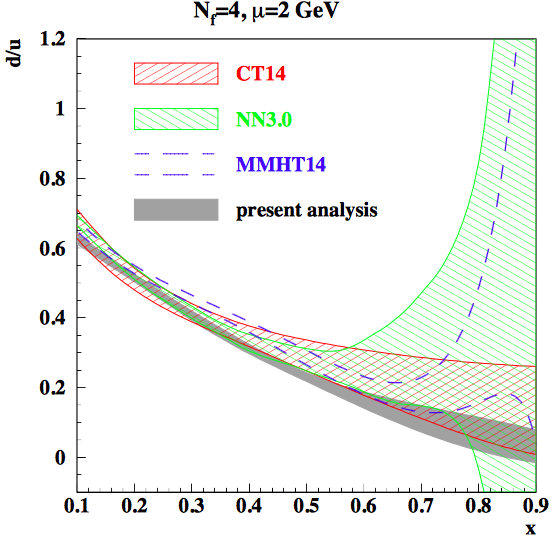}
\caption{
Ratio of down to up valence PDF for various PDF sets.  The label 'present analysis' refers to \cite{Alekhin:2015cza}.
}
\label{fig:alekhin}
\end{center}
\end{figure}
There is an open question on the form of $\bar d -\bar u$ at small $x$
and $d/u$ as $x \to 1$. Strong claims about $\bar d -\bar u$ at small $x$ 
being non-zero and $d/u$ as $x \to 1$ being very well constrained (see Fig.~\ref{fig:alekhin}) were 
made based on fits~\cite{Alekhin:2015cza,Alekhin:2016uxn} 
including $W^+$ asymmetry data from $D0$ and
$W^+, W^-$ data from 
LHCb. However, the CJ15 fit \cite{Accardi:2016qay}
includes $D0$ data 
\cite{Abazov:2013dsa,D0:2014kma} 
with a good fit quality 
(as do CT14, but with a poorer fit),
and see no requirement for any small-$x$ $\bar d -\bar u$ difference.  
Similarly, MMHT14 compares very well~\cite{Thorne} to
data on high rapidity $W$ production at LHCb at
$7~\TeV$~\cite{Aaij:2015gna} 
and many PDF sets give predictions which compare well to $W$ production 
at  LHCb at
$8~\TeV$~\cite{Aaij:2015zlq}.
Fits to more precise vector boson data is 
an obvious area for further study.

A number of talks addressed the  
flavour decomposition of the proton from another direction.
One of the measures of  $\bar d -\bar u$ is the Gottfried Sum Rule. 
It was described how one can effectively measure free neutron PDFs using
the BONUS (Barely Off-shell Neutron Structure) experiment 
\cite{Griffioen:2015hxa}, where in electron deuteron scattering one measures
the scattered electron in coincidence with a  proton tag. This study 
allows for a re-examination of the 
Gottfried sum rule from NMC deuteron scattering data \cite{Arneodo:1994sh}
down to $x=0.004$ (relying on an extrapolation at higher $Q^2$).
There is no sign in this of  $\bar d -\bar u$ changing sign or being large 
at low $x$. It was also shown how by performing a simultaneous study of 
precision proton and deuteron data one can fit and also verify predictions 
for deuteron corrections \cite{Accardi:2016qay}, e.g. DO $W$ asymmetry 
data and deuteron DIS probe the down quark for similar $x$ so a simultaneous 
fit largely determines the deuteron correction.

\begin{figure}[h]
\begin{center}
\includegraphics[width=15cm]{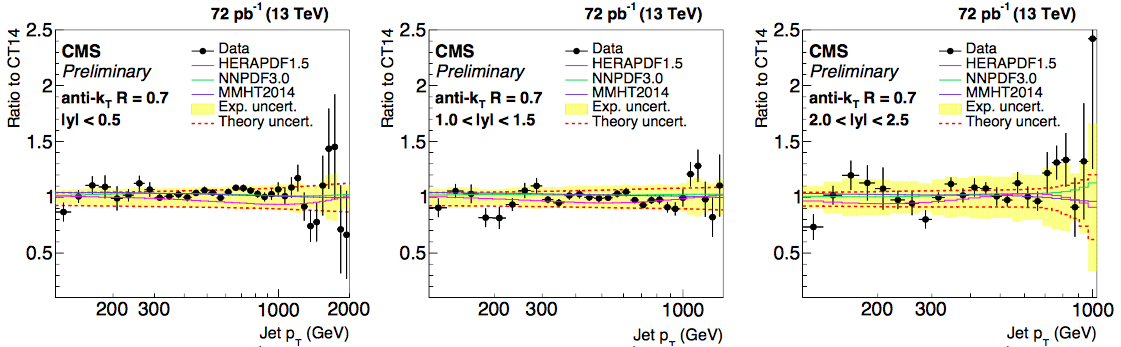}
\caption{Jet cross-section ratios to ME calculations using CT14 PDFs.  The effect of other PDFs is shown by different coloured lines~\cite{cms}.}
\label{fig:cms}
\end{center}
\end{figure}

\section{Experimental inputs to PDFs}

\begin{figure}
\begin{center}
\includegraphics[width=15cm]{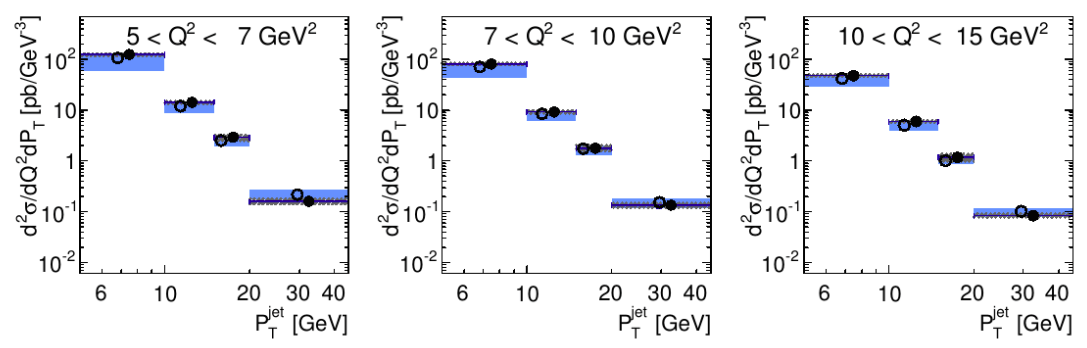}
\caption{Preliminary results from H1 on inclusive jet cross-section (solid points).  The open points are a previous analysis that just used HERA-I data.  The coloured band represents the NLO predictions~\cite{hjet}.}
\label{fig:hjet}
\end{center}
\end{figure}

Several new experimental results that will help to further constrain the PDFs were shown.
The CMS collaboration presented inclusive jet cross-sections at 13 TeV~\cite{cms}.  Fig.~\ref{fig:cms}
shows their data compared to CT14 and various other PDF sets.
The other end of the energy spectrum was represented by results from H1~\cite{hjet} who showed inclusive
jet cross-sections at low-$Q^2$, some of which are shown in Fig.~\ref{fig:hjet}.  
The experimental uncertainties are dominated by the jet energy scale and modelling and are
significantly smaller than the theory uncertainties, which have large scale uncertainties.

\begin{figure}
\begin{center}
\includegraphics[width=7cm]{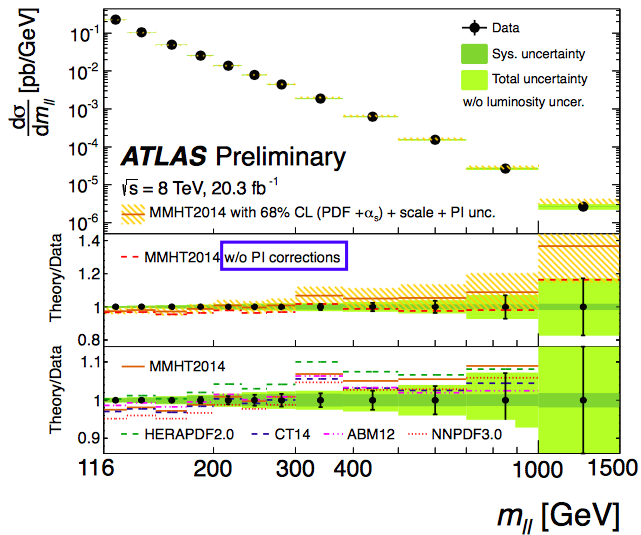}
\includegraphics[width=8cm]{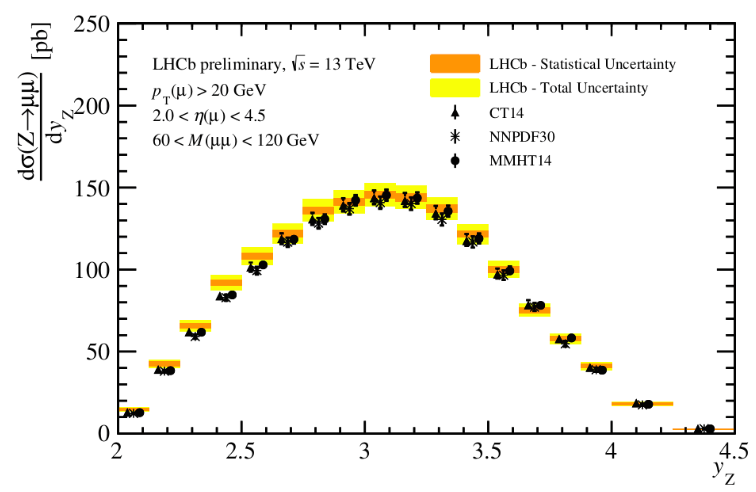}
\caption{Left: ATLAS results on Drell-Yan production at 8 TeV~\cite{atlas13}.  Right: LHCb results on Z boson production at 13 TeV~\cite{lhcb13}.  Both are compared to predictions with various PDF sets.}
\label{fig:atlaslhcb}
\end{center}
\end{figure}

The ATLAS collaboration presented results on Drell-Yan production including a
recent analysis of 8 TeV data~\cite{Aad:2016zzw}.
Fig.~\ref{fig:atlaslhcb} (left) shows the cross-section for electron-positron pairs with masses
between 116 and 1500 GeV, compared to different PDFs.
It was demonstrated that these data have sensitivity 
to the poorly known photons PDFs and can thus be used to constrain them.

Results on W boson production at 13 TeV were shown by ATLAS~\cite{atlas13}, while Z boson production
at 13 TeV was shown by the ATLAS and LHCb~\cite{lhcb13} collaborations.  Fig.~\ref{fig:atlaslhcb} (right) shows
the differential cross-section for Z production in the forward region as measured by LHCb, compared
to predictions using various PDFs.

\section{Nuclear PDFs}

\begin{figure}
\begin{center}
\includegraphics[width=6.5cm]{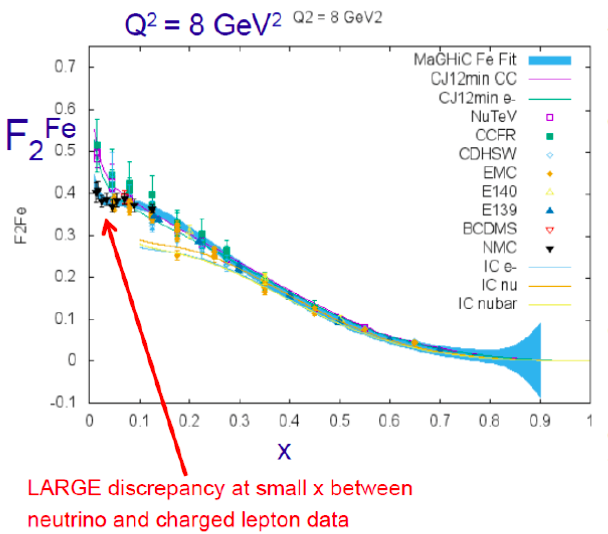}
\includegraphics[width=8cm]{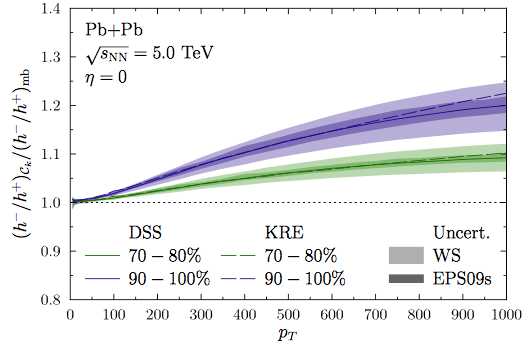}
\caption{Left: Comparison of $F_2^{Fe}$ measured in neutral and charged current data~\cite{Keppel}.
Right: Ratio of charged hadrons normalised with minimum bias events for 
two fragmentation functions.  No skin effect corresponds to a value of one~\cite{Helenius:2016dsk}.
}
\label{fig:nucl}
\end{center}
\end{figure}

There were various presentations of nuclear PDFs in the light of new LHC data.
The updated nCTEQ15 PDFs \cite{Kusina:2016lxo}  
include the impact of some new LHC data on $p-Pb$ scattering via the
reweighting technique, but this provides little impact on the PDFs yet. 
No neutrino data is included in this fit. There was also an 
update on Kulagin-Petti nuclear PDFs (see e.g. \cite{Kulagin:2015lkm,
Ru:2016wfx}), which 
model the nuclear corrections rather than primarily fit them to data. 
This includes account of Fermi motion, off-shell effects, nuclear meson 
exchange current corrections and contributions from coherent nuclear 
interactions (nuclear shadowing). So far predictions for LHC $p-Pb$ data
appear to be successful.
The difference between neutral and charged current nuclear
data was investigated for iron targets \cite{Keppel}. 
As seen in Fig.~\ref{fig:nucl} (left),
very good agreement between the two is found at high $x$, but there is 
evidence for less suppression in the
neutrino nuclear structure functions at low $x$. There were also 
a couple of detailed studies in 
nuclear collisions at the LHC. For example, 
the centrality dependence of nuclear modifications was investigated 
\cite{Kohler} by looking at $W$ production as a function of rapidity
in three centrality bins, with some evidence for centrality dependence seen. 
Another topic is the so called neutron skin effect. The neutron distribution 
is expected to be broader than that of the proton in the nucleus
\cite{Helenius:2016dsk}, i.e. the neutron tail extends further. This can 
potentially be seen by looking at the charged hadron ratio
in $Pb-Pb$ collisions
where deviations from unity are expected with increasing 
$p_T$ and rapidity.
Fig.~\ref{fig:nucl} (right) shows the effect at mid-rapidity: at forward rapidities, deviatinos set in at lower $p_T$,
although the uncertainties are larger.

We also saw a presentation of the angular distributions of Drell-Yan dimuons at E906/SeaQuest testing the correlation of the azimuthal and polar angles of leptonic products relative to the initial hadronic plane\cite{Ramson}, and a study of the ability of pion exchange models to describe both leading neutron electroproduction at HERA and to extract the $\bar d\ne \bar u$ flavour asymmetry in the proton \cite{McKenney:2015xis}.

\section{Transverse Momentum Dependent (TMD) PDFs}

The session devoted to TMD PDFs highlighted the importance of including resummations
and nonperturbative information, particularly at low $k_t$ , and a proper simulation of 
parton showers. It included various updates on a new large scale and ambitious project 
including full coupled quark and gluon evolution using the Monte Carlo approach in a 
form applicable over all $x$ and $Q^2$. Results on the fully integrated PDFs obtained 
in the framework were successfully compared to HERA inclusive cross section data, and 
updates on the considerable ongoing work for the fully unintegrated PDFs were 
presented~\cite{Lelek}. Saturation effects in the same framework were 
discussed~\cite{Monafred}, and TMDlib libraries and a plotter were introduced for 
the first time~\cite{Connor}, also produced by the same group. There was also a 
contribution studying the role of the nonperturbative input to the (TMD) gluon density 
in hard processes at the LHC, and deriving the input distribution from a fit of 
inclusive hadron spectra measured at low transverse momenta in pp collisions at the 
LHC \cite{Grinyuk:2015lna}.

\section{Theoretical topics}

\begin{figure}
\begin{center}
\includegraphics[width=7cm]{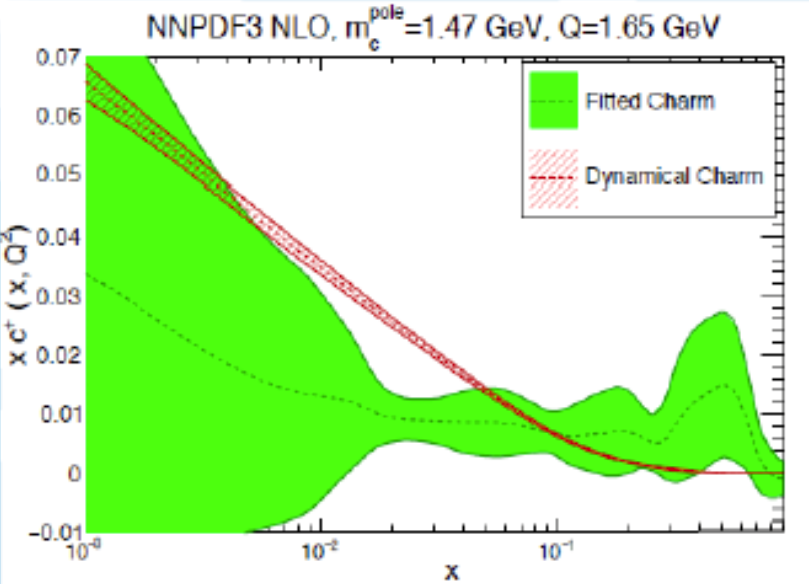}
\includegraphics[width=7cm]{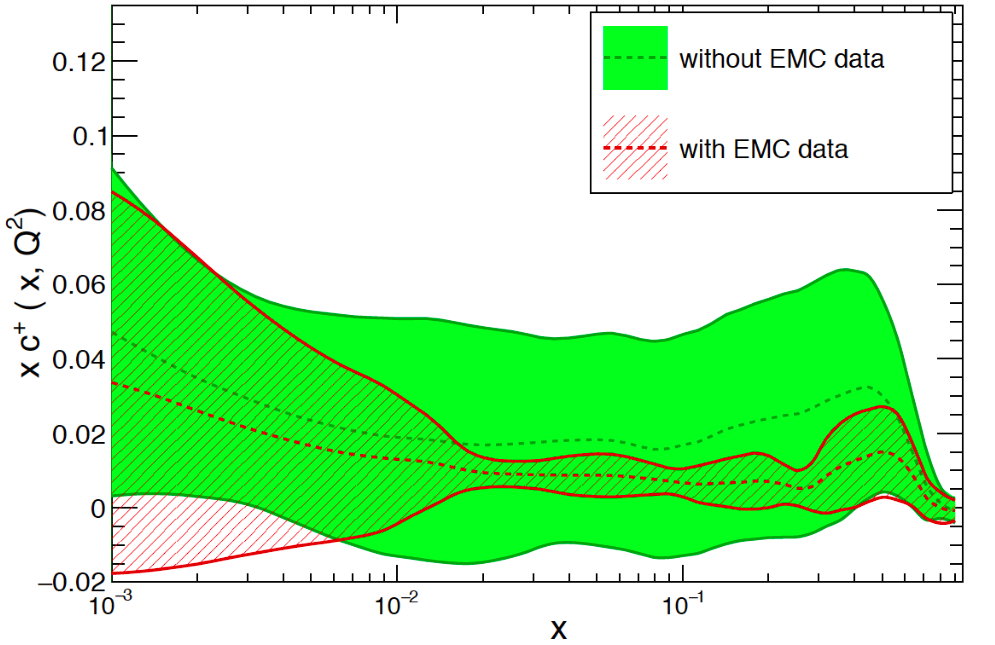}
\caption{NNPDF fitted charm compared to dynamical charm (left) and to the fit without 
EMC data (right)~\cite{Ball:2016neh}.
}
\label{fig:charm}
\end{center}
\end{figure}

There were a variety of updates on specific theoretical 
topics relevant for PDF studies. There was a presentation of intrinsic charm 
based on ~\cite{Rostami:2015iva}, which looked at the impact on the 
differential $\gamma +c$ cross section, showing this could be significant. 
Intrinsic charm was also discussed as 
part of a NNPDF study \cite{Ball:2016neh}, where it is
determined from a fit to heavy flavour distributions, on top of 
that generated dynamically from the gluon and light quarks via evolution.
The inclusion of EMC data~\cite{Aubert:1982tt} 
significantly reduces the 
uncertainties and the fitted charm is lower than purely dynamical charm for $x\sim 0.05$
(see Fig.~\ref{fig:charm}).
This has implications for predictions of $W,Z+c$ data at the LHC.

The NNPDF group also presented PDFs with threshold resummations 
\cite{Bonvini:2015ira}, whose effects are much larger at NLO 
than NNLO, as NNLO already includes much of the effect present at NLO. 
Data sets for which the threshold corrections are unknown are not used in the fit.
The resummed cross-sections are closer to the fixed order predictions, although
the effect at NNLO is arguably no bigger than that due to the missing data. 

Another study concentrated on the extreme limits of the PDFs~\cite{Ball:2016spl}, 
comparing the effectively small-$x$ powers and
high-$x$ powers of $(1-x)$ for different PDF sets, finding some significant 
variation in the latter for the gluon. 
Improvements to the low-$x$ gluon for NNPDF were shown~\cite{rottoli}, obtained by incorporating charm
production data from LHCb.  This allows improved predictions for the prompt 
atmospheric neutrino flux up to $10^8$ GeV, which are consistent with IceCube bounds.

There was an investigation of different methods to incorporate the effect of photons in hard processes \cite{Luszczak:2015aoa}. Two different approaches were used for calculating cross sections: either the photon is treated as a collinear parton in the proton, or alternatively the $k_T$ factorization method is used. Also we had a discussion of a method to obtain the double gluon distribution from the single gluon distribution using sum rules \cite{Golec-Biernat:2016qtk}.

There was s discussion of exclusive $J/\Psi$ and $\Upsilon$ production 
\cite{Jones:2015nna}, which in principle can constrain the gluon PDF
that is related to generalised parton distributions.
The one-loop corrections to the cross section have been calculated but 
lead to large, opposite sign corrections, particularly for $J/\Psi$. 
Theoretical improvements are needed
for these processes to be a precision constraint on the gluon 
distribution. 

A procedure for calculating PDFs on the Lattice~\cite{Alexandrou:2015rja}
was presented, using quasi-distributions 
at finite longitudinal momentum rather than exact distributions 
in the infinite momentum frame.
One can then try to transform the quasi-distributions to the correct limit. 
Longitudinal momentum $P_3=2\pi/L$ on available lattices 
corresponds to about $~0.4~\GeV$. The values used are hence limited 
to about $1.6~\GeV$ at best at present, but one can see improvements
in results as better lattices are used. We look forward to future 
developments in this alternative method of PDF determination.

\section{Tools for PDFs}
There were a number of updates on various tools used to study and 
present PDFs. 
APFEL~\cite{Bertone:2016lga} showed a plotting procedure for 
changing the transition point for heavy flavours in variable flavour schemes,
which demonstrates diminished sensitivity to this in PDFs at NNLO 
compared to NLO. 
The HERAFitter collaboration, now called XFitter \cite{::2016pbr}, 
presented theoretical improvements and the inclusion of new data sets.
The effect of asymmetric uncertainties was discussed in the
CT14 replica PDFs presentation~\cite{Hou:2016sho}, along 
was the $\chi^2$ distribution of the global fit quality for 1000 
CT14 replicas,
which use 28 PDF eigenvectors and tolerance $\sim 40$ for one sigma.
Remarkably, this is extremely similar to the equivalent $\chi^2$ distributions 
for 1000 NNPDF replicas (see e.g. \cite{Ball:2011mu}), despite the 
fact these are obtained from a very 
different approach. This supports the use of a tolerance criterion in global 
fits based on the eigenvector approach.

\end{document}